\documentclass[aps,prl,twocolumn,showpacs]{revtex4}

\usepackage[dvips]{graphicx}
\usepackage{dcolumn}
\usepackage{bm}
\begin{document}

\title{Localized Distributions of Quasi Two-Dimensional Electronic States near Defects Artificially Created at Graphite Surfaces in Magnetic Fields}

\author{Y. Niimi}
\email{yasuhiro.niimi@grenoble.cnrs.fr}
\altaffiliation{Present address: Institut Neel, CNRS, B.P. 166, 38042 Grenoble Cedex 09, France}
\author{H. Kambara}
\author{Hiroshi Fukuyama}
\affiliation{Department of Physics, University of Tokyo, 7-3-1 Hongo Bunkyo-ku, Tokyo 113-0033, Japan}

\date{January 13, 2009}

\begin{abstract}
We measured the local density of states of 
a quasi two-dimensional electron system (2DES)
near defects, artificially created by 
Ar-ion sputtering, on surfaces of highly oriented 
pyrolytic graphite (HOPG) with scanning tunneling spectroscopy (STS) 
in high magnetic fields. 
At valley energies of the Landau level spectrum, we found 
two typical localized distributions of the 2DES depending on the defects. 
These are new types of distributions which are not observed in the previous 
STS work at the HOPG surface near a point defect 
[Y. Niimi \textit{et al}., Phys. Rev. Lett. {\bf 97}, 236804 (2006).]. 
With increasing energy, we observed gradual transformation from 
the localized distributions to the extended ones 
as expected for the integer quantum Hall state. 
We show that the defect potential depth is responsible for 
the two localized distributions from comparison with theoretical calculations. 
\end{abstract}

\maketitle

Two-dimensional electron systems (2DESs) exhibit 
fascinating quantum phenomena at low temperature. 
The quantum Hall (QH) effect in high magnetic fields and 
the Anderson localization are two well-known examples~\cite{yoshioka1}. 
These phenomena have been studied mainly by means of transport measurements. 
However, it is generally difficult to investigate the electronic states of 
the 2DESs in real space on nanometer scale with local probes, 
since they are usually formed at heterojunctions a few hundreds nanometers 
below the semiconductor surfaces. 

Recently, 2DESs formed 
at semiconductor surfaces~\cite{kanisawa1,morgenstern,ono,suzuki} 
and semimetal ones~\cite{matsui,niimi1} 
have been studied with scanning tunneling microscopy 
and spectroscopy (STM/STS). 
These are powerful techniques to investigate 
the local density of states (LDOS) at sample surfaces. 
Morgenstern \textit{et al}.~\cite{morgenstern} observed 
clear Landau quantization as well as complicated patterns of 
the LDOS depending on bias voltage for a two-dimensional electron gas (2DEG)
at a cleaved InAs(110) surface 
with submonolayer iron deposition. 
However, the localized and extended LDOS 
distributions are not clearly distinguished in their measurements. 

More distinct localization and extension of the LDOS depending on energy, 
which indicate the possible QH state, have been observed near a point defect 
at surfaces of highly oriented pyrolytic graphite (HOPG)~\cite{niimi1}. 
At valley energies of the Landau levels (LLs), 
a circular distribution of the LDOS was observed near the defect. 
The distribution with a radius comparable to the magnetic length 
$l_{B}$ ($=\sqrt{\hbar/eB}$ where $B$ is magnetic field) 
was semiquantitatively explained by the calculated LDOS 
for 2DEG in magnetic fields in the $1/r$ potential~\cite{niimi1,yoshioka2}. 
In transport measurements for HOPG~\cite{kopelevich}, 
a Hall resistance plateau was observed, 
which is also indicative of the QH effect in this material. 
Unambiguous QH plateaus were observed in the 2DESs 
at a single layer of graphite (graphene)~\cite{geim1,kim1} 
and bilayer graphene~\cite{geim1}. 
Since these thin-layer graphite systems including HOPG 
have the surface 2DEGs, 
they provide ideal arenas for investigation of quantum phenomena in 2DESs, 
especially the QH state, with the STM/STS techniques. 

In this Letter, we studied the LDOS at HOPG surfaces 
with randomly distributed defects which are 
artificially created by Ar-ion sputtering. 
Two types of the localized LDOS distributions are observed 
at the valley energies of the LLs 
in the differential tunnel conductance (d$I/$d$V$) images. 
As we increase energy, the distributions are extended to follow 
the complicated potential landscapes. 
We reveal that the functional forms of the defect potentials are 
closely related to the LDOS distributions. 

The STM/STS measurements were performed 
at temperatures below 30 mK and in magnetic fields up 
to 6 T using an ultra low temperature STM~\cite{ult-stm}. 
Electrochemically etched W wires were used as STM tips. 
The d$I/$d$V$ curves and images were 
taken by the lock-in technique with a bias modulation
$V_{\rm mod}$ of 0.5 or 1.0 mV at a frequency of 412 Hz.
The HOPG sample~\cite{hopg} was cleaved in air and then quickly loaded 
into an ultra high vacuum chamber of the STM. 
Defects were made by sputtering the sample surface \textit{in situ} 
with a 30 eV Ar-ion beam for a few seconds 
in an Ar atmosphere of $10^{-6}$ Pa. 
Higher energy beams for a longer time made the surfaces too rough 
for the STM/STS measurements. 

Figure 1(a) shows a typical STM image over $100 \times 100$ nm$^{2}$ 
at the sputtered HOPG surface. 
16 defects (Defects 1 $\sim$ 16) are seen in this image. 
The averaged defect density is $2 \times 10^{11}$ cm$^{-2}$. 
The defects have nearly the same diameters ($6\sim 8$ nm) which are much 
larger than that of the point defect ($\leq 1$ nm) studied 
in Ref.~\cite{niimi1}. 
They are randomly distributed on the surface. 
In Fig. 1(b), we show close-up images of two of them, where 
complicated inside structures are clearly seen. Just outside the defects, 
the $(\sqrt{3}\times \sqrt{3})R30^{\circ}$ superstructure and honeycomb one 
consisting of the B-site carbon atoms are observed~\cite{superlattice,niimi2}. 

\begin{figure}
\begin{center}
\includegraphics[width=8.5cm]{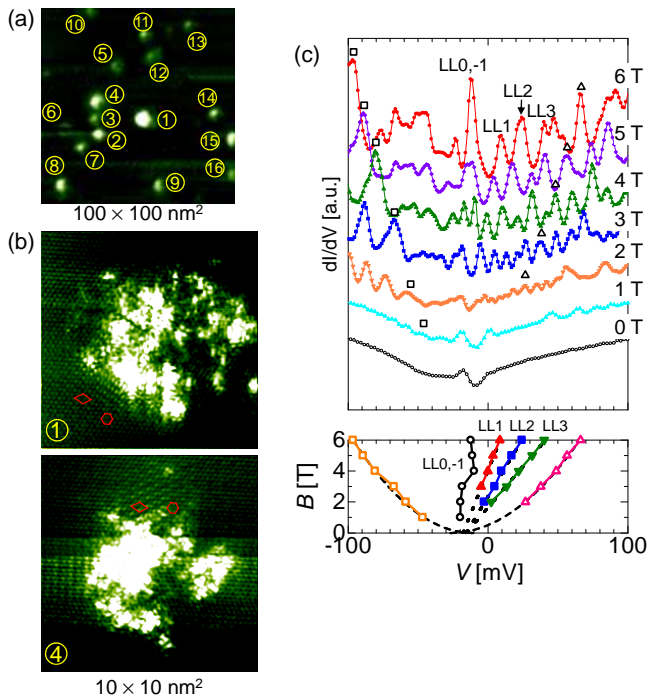}
\caption{(a) Typical STM image at the sputtered HOPG surface [$100 \times 100$ nm$^{2}$, $V=120$ mV, $I=0.2$ nA]. The numbers are assigned for all (sixteen) defects in this image. (b) Close-up STM images [$10 \times 10$ nm$^{2}$] of Defects 1 and 4. The diamond and honeycomb represent the $(\sqrt{3}\times \sqrt{3})R30^{\circ}$ superstructure and honeycomb one. (c) Top figure shows tunnel spectra at the sputtered surface in several different magnetic fields perpendicular to the graphite basal plane. Each spectrum is shifted for clarity. Bottom figure shows the LL peak energies as a function of $B$. The dotted and broken lines are linear and square root fittings of the experimental data, respectively.}
\label{fig1}
\end{center}
\end{figure}

In Fig. 1(c), we show tunnel spectra at several different magnetic fields 
averaged over the whole scan area of Fig. 1(a). 
Clear LL peaks similar to those in the previous works~\cite{matsui,niimi1} 
are observed. 
The Landau index ($n$) for each peak is assigned as in Ref.~\cite{niimi1}. 
By comparing with the calculated surface LDOS, 
the effective thickness of this area is estimated as 
about 20 layers ($=6.7$ nm) of graphene~\cite{matsui}. 
Note that the Fermi energy ($E_{F}$) here is shifted 
by $+30$ meV, compared to those 
in the previous works~\cite{matsui,niimi1}. 
This is because a large number of defects made by sputtering 
inclined the electron and hole balance to excess electrons. 

It is intriguing to recognize that the LLs have two different 
magnetic field variations depending on $n$. 
For example, the peak energies of LL1, LL2 and LL3 near $E_{F}$
are proportional to $B$ as expected for the conventional 2DEG. 
Meanwhile, the LLs far from $E_{F}$ denoted by open symbols 
in Fig. 1(c) have a $\sqrt{B}$ dependence. 
If we fit the latter LLs with $E_{n}=\sqrt{2e\hbar v_{F}^{2} B|n|}$, 
the formula of the LL energies 
for \textit{massless} Dirac Fermions~\cite{note}, 
we obtain the Fermi velocity $v_{F}$ of $1\times 10^{6}$ m/s. 
This value is consistent with the estimation from 
the transport measurements of graphene~\cite{geim1,kim1}. 
These facts, 
which are consistent with similar results of Ref.~\cite{sts_hopg}, 
indicate that there are 
linear and parabolic subbands 
at the HOPG surface~\cite{band_dispersion}. 

\begin{figure*}
\begin{center}
\includegraphics[width=13.8cm]{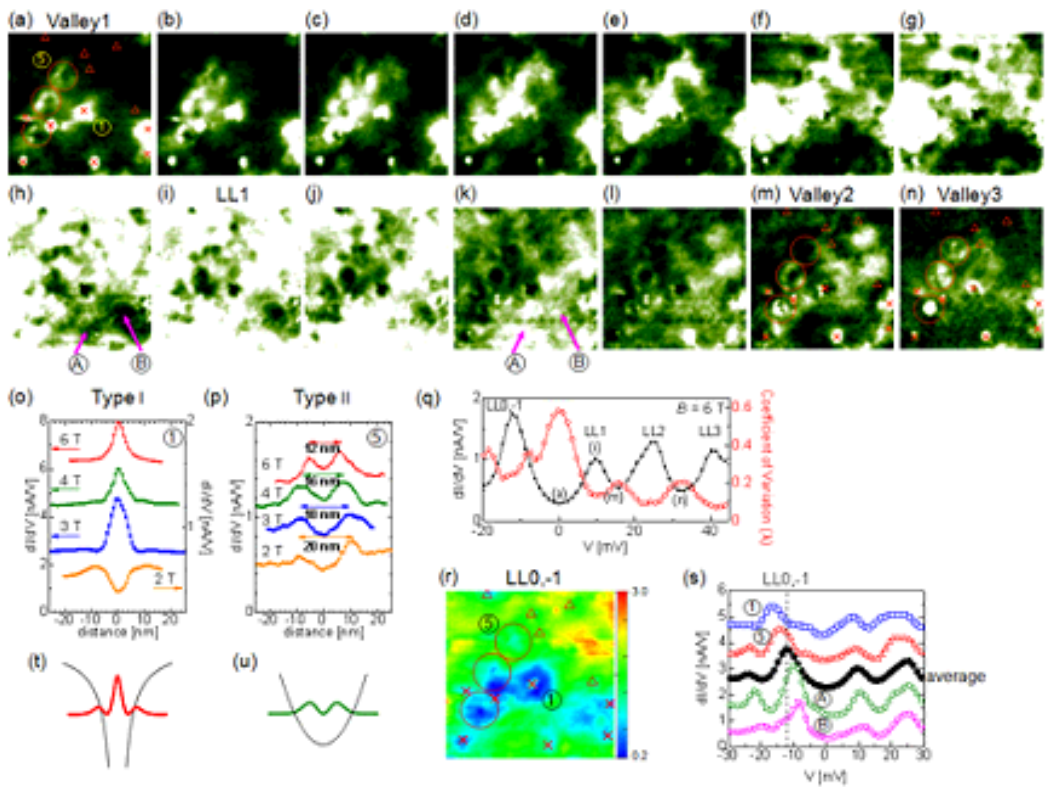}
\caption{(a)-(n) d$I/$d$V$ images in the same scan area as Fig. 1(a) at various bias voltages [$100 \times 100$ nm$^{2}$, $B=6$ T, $T=30$ mK, $I=0.2$ nA, $V_{\rm mod}=1.0$ mV]; (a) $-0.1$, (b) 1.1, (c) 2.3, (d) 3.5, (e) 4.7, (f) 5.9, (g) 7.1, (h) 8.3, (i) 9.5, (j) 10.7, (k) 11.9, (l) 13.1, (m) 14.3, (n) 31.6 mV. The defect positions are denoted by the crosses, solid circles and triangles, depending on the kind of the localized LDOS distribution (see text). The arrows in (h) and (k) show the regions (Regions A and B) where the electronic states are hardly extended with increasing energy. (o), (p) Cross sections of the d$I/$d$V$ images averaged radially near (o) Defect 1 and (p) Defect 5 at the valley energies in between LL0,$-$1 and LL1 at different fields. Each cross section is shifted by (o) 2 or (p) 0.4 nA/V for clarity. The arrows in (p) show the approximate diameters of the LDOS rings. (q) Average (solid circle) and coefficient of variation (open circle) of d$I/$d$V$ in the whole scan area as a function of bias voltage. (r) d$I/$d$V$ image at the peak energy of LL0,$-$1 ($-$12 mV). (s) Tunnel spectra averaged over the whole scan area (black), over $10 \times 10$ nm$^{2}$ near Defect 1 (blue), Defect 5 (red), Region A (green), and Region B (pink). Each spectrum is shifted by 1 nA/V for clarity. (t), (u) Schematics of the wave functions for the ground states in (t) the $1/r$ and (u) harmonic potentials. }
\label{fig2}
\end{center}
\end{figure*}

Next, in Figs. 2(a)-2(n), we show d$I/$d$V$ images 
at different bias voltages at 6 T in the same scan area as Fig. 1(a). 
Note that the same contrast is used for all the images. 
At the valley energy in between LL$0,-1$ and LL1 [Fig. 2(a)], 
the electronic states are strongly or weakly localized 
near some of the defects. 
The localization is gradually dispersed with increasing energy 
until LL1 [Figs. 2(b)-2(i)]. 

First of all, we focus on the d$I/$d$V$ image at the valley energy 
in Fig. 2(a). 
We found two new types of localized LDOS distributions 
depending on the defects, which are not observed in the 
previous STS work~\cite{niimi1}.
One is a distribution where the LDOS amplitude has a maximum just on defect 
(Type I: denoted by cross). 
The other is a ring distribution around defect 
(Type II: denoted by solid circle). 
We also found that five defects do not support any localized states nearby
(denoted by triangle). 
In Figs. 2(o) and 2(p), we show cross sections of Type I (Defect 1) 
and Type II (Defect 5) at different fields. 
Both distributions have diameters of the order of 
$l_{B}$ ($\propto 1/\sqrt{B}$) at each field, but 
the peak amplitude of Type I is much larger 
than that of Type II. 
The latter is qualitatively different from the 
LDOS distribution observed near the point defect~\cite{niimi1}, while
the former is seemingly similar to it. 
However, Type I does not have an appreciable satellite ring 
unlike the point defect case. 
And it suddenly changes to Type II-like in between 2 and 3 T 
with decreasing field. 
Such behavior is also observed for Defects 2, 6, and 16 
but not for Defects 8 and 9. 

As we increase energy [Figs. 2(b)-2(i)], 
the LDOS localization is gradually dispersed, 
showing complicated extended patterns. 
It should be noted that the electronic states are extended to keep away, 
for instance, regions denoted by arrows 
(Regions A and B in Fig. 2(h)). 
In contrast, as energy exceeds LL1 by 2 meV [Fig. 2(k)], 
the electronic states are existing in those regions, 
and also Type II distributions can be seen 
near some of the defects (center-left). 
A similar evolution is observed at energies in between other two 
successive LLs below LL3 [Figs. 2(m) and 2(n)]. 

The localized and extended states can be discriminated more quantitatively 
by plotting the coefficient of variation ($\lambda$)
for each d$I/$d$V$ image 
(which is given by dividing the standard deviation by the average) 
as a function of bias voltage in Fig. 2(q). 
It is obvious from this plot that 
the spatial variations of LDOS are larger 
at the valley energies in 
between two LLs, while they are smaller at 
the peak energies of the LLs. 
This is an additional indication for a transition 
from localized to extended states. 

Figure 2(r) is the d$I/$d$V$ image at the peak energy of LL$0,-1$. 
As is shown in the previous STS works on graphite~\cite{matsui,niimi1},
this peak is much more sensitive to 
the local electrostatic potential compared to the other LL peaks. 
This is attributable to much narrower but 
dispersive subband structure of LL$0,-1$
along the $k_{z}$ axis [see Fig. 2(b) of Ref~\cite{matsui}].
Thus, we assume that the image at the LL$0,-1$ peak energy 
is indicative of the surface potential distribution 
which should be crucial to the electron localization. 
For instance, there are not strong potential variations around the 
defects which do not support any localized states in Fig. 2 (a).
Figure 2 (s) shows the tunnel spectra near Defect 1 (Type I) 
and Defect 5 (Type II). 
The LL$0,-1$ peak energies near Defect 1 ($V=-18$ mV) and Defect 5 ($-15$ mV) 
are shifted to the negative bias voltage side 
compared to that averaged over the whole scan area ($-12$ mV). 
This indicates that the potential of Defect 1 is deeper 
than Defect 5 and that both are confining ones. 
The same tendencies are seen for the other defects for Types I and II.
On the other hand, in Regions A and B, the LL$0,-1$ peak energies are shifted 
to the positive energy side [Fig. 2(s)], which indicates 
that the potentials of those regions 
are relatively high compared to the other regions. 
These results suggest 
that the observed energy and spatial variations of the LDOS 
are determined by the surface potential distribution.
This is expected behavior for the localized and extended states 
in the QH effect~\cite{ando}.

The two different localized LDOS distributions, i.e., Types I and II, 
can be qualitatively explained by recent calculations 
on eigenfunctions of simple 2DEG in magnetic fields~\cite{yoshioka2}.
With a spatial variation of electrostatic potential, 
the degeneracy of the LLs is lifted. 
In the case of $1/r$ potential, 
the angular momentum ($L_{z}=\ell_{z}\hbar$) 
of the ground state is zero for each LL.
Thus, its wave function has a maximum at the origin 
to minimize the potential energy [Fig. 2(t)]. 
On the other hand, the ground state in the harmonic potential has 
$\ell_{z}=n$, and the maximum of electron probability forms a 
ring around the origin [Fig. 2(u)]. 
The distributions of Types I and II are basically consistent with 
the ground state ones calculated 
for the $1/r$ and harmonic potentials, respectively. 

There are, however, a few experimental observations which cannot be described 
well by theoretical calculations. 
For example, the LDOS satellite ring for Type I is less obvious than 
that calculated for the $1/r$ potential. 
This can be alleviated with shallower potentials such as 
$1/\sqrt{r^{2}+d^{2}}$ ($d$ is the depth from the surface). 
The localized LDOS distribution changes 
from Type I to II below 3 T, 
while the calculated LDOS has Type I like distribution down to 
much lower field~\cite{yoshioka2}. 

\begin{figure}
\begin{center}
\includegraphics[width=6cm]{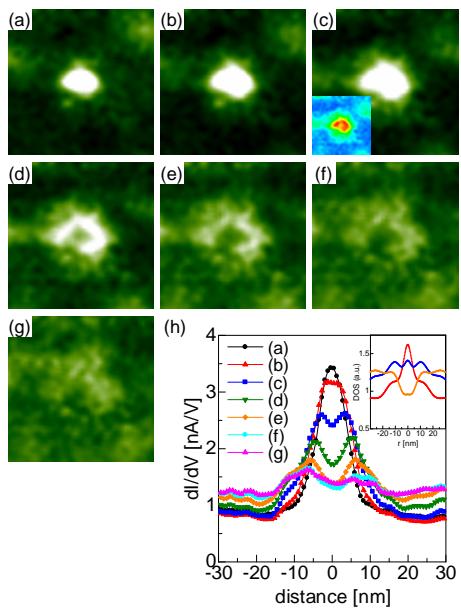}
\caption{(a)-(g) d$I/$d$V$ images near a defect at various bias voltages [$50 \times 50$ nm$^{2}$, $B=6$ T, $T=30$ mK, $I=0.2$ nA, $V_{\rm mod}=0.5$ mV]; (a) 29, (b) 30, (c) 32, (d) 33, (e) 34, (f) 35, (g) 37 mV. The same contrast is used for all the d$I/$d$V$ images. The inset of figure (c) shows the same d$I/$d$V$ image in color scale. (h) Cross sections of the d$I/$d$V$ images averaged radially near the defect. The inset is calculated LDOS at different energies in the $1/\sqrt{r^2+d^2}$ potential shown in Fig. 11 of Ref. [8].}
\label{fig3}
\end{center}
\end{figure}

Finally, let us describe a detailed energy dependence 
of Type I localized distribution in Fig. 3. 
This observation became possible near a similarly created but isolated defect 
at an HOPG surface with a less defect density. 
At the valley energy in between LL1 and LL2, the electronic state 
is localized just on the defect. 
With increasing energy, however, the LDOS maximum is 
shifted from the center to the outside, and forms 
the ring-type distribution [Figs. 3(a)-3(c)]. 
The LDOS ring gradually expands and disappears at 
the peak energy of LL2 [Figs. 3(d)-3(g)]. 
This evolution is sumarized in Fig. 3(h) 
where cross sections of the images are plotted. 
Similar evolutions are also but less obviously observed in the LDOS 
near other defects shown in Figs. 2(a)-2(i). 
We don't have thorough theoretical explanations 
for this observation at the moment. 
A qualitatively similar energy dependence of localized state 
has been given in the previous calculations 
with the $1/\sqrt{r^2+d^2}$ potential [inset of Fig. 3(h)]. 
The essential point here is the spectral weight conversion 
from the ground state to the higher-order excited states. 
The improvement of the caluculations by tuning the functional form of 
potential is highly desirable. 

In conclusion, we have measured the LDOS near defects artificially created by 
Ar-ion sputtering at HOPG surfaces 
in high magnetic fields with STM/STS. 
At the valley energies in between the adjacent LLs, 
the two new localized distributions were observed 
depending on the kinds of the defect potentials.  
With increasing energy, the localized distributions are extended to follow the 
potential landscapes created by the randomly distributed defects 
as is expected for the QH system. 

We thank D. Yoshioka for valuable discussions. 
This work was financially supported by Grant-in-Aid for Scientific Research 
on Priority Areas (Grant No. 17071002) from MEXT and ERATO Project of JST. 
Y.N. acknowledges the JSPS Research program for Young Scientists. 

\textit{note added}: After submitting, we became aware of 
a related STM/STS work on a semiconductor surface~\cite{hashimoto} showing similar results to ours.


\begin{thebibliography}{00}
\bibitem{yoshioka1}
See, for example, D. Yoshioka, \textit{The Quantum Hall Effect} (Springer, Berlin, 2002).
\bibitem{kanisawa1}
K. Kanisawa \textit{et al}., Phys. Rev. Lett. {\bf 86}, 3384 (2001).
\bibitem{morgenstern}
M. Morgenstern \textit{et al}., Phys. Rev. Lett. {\bf 90}, 056804 (2003).
\bibitem{ono}
M. Ono \textit{et al}., Phys. Rev. Lett. {\bf 96}, 016801 (2006).
\bibitem{suzuki}
K. Suzuki \textit{et al}., Phys. Rev. Lett. {\bf 98}, 136802 (2007).
\bibitem{matsui}
T. Matsui \textit{et al}., Phys. Rev. Lett. {\bf 94}, 226403 (2005).
\bibitem{niimi1}
Y. Niimi \textit{et al}., Phys. Rev. Lett. {\bf 97}, 236804 (2006).
\bibitem{yoshioka2}
D. Yoshioka, J. Phys. Soc. Jpn. {\bf 76}, 024718 (2007).
\bibitem{kopelevich}
Y. Kopelevich \textit{et al}., Phys. Rev. Lett. {\bf 90}, 156402 (2003).
\bibitem{geim1}
K. S. Novoselov \textit{et al}., Nature (London) {\bf 438}, 197 (2005).
\bibitem{kim1}
Y. B. Zhang \textit{et al}., Nature (London) {\bf 438}, 201 (2005).
\bibitem{ult-stm}
H. Kambara \textit{et al}., Rev. Sci. Instrum. {\bf 78}, 073703 (2007).
\bibitem{hopg}
Super Graphite (grade MB), Matsushita Electric Industrial Co., Ltd. 
\bibitem{superlattice}
P. Ruffieux \textit{et al}., Phys. Rev. B {\bf 71}, 153403 (2005).
\bibitem{niimi2}
Y. Niimi \textit{et al}., Appl. Surf. Sci. {\bf 241}, 43 (2005); 
Y. Niimi \textit{et al}., Phys. Rev. B {\bf 73}, 085421 (2006).
\bibitem{note}
In this fitting, we assume that the LL$0,-1$ for HOPG corresponds to LL0 for graphene. 
\bibitem{sts_hopg}
G. Li and E. Y. Andrei, Nature Phys. {\bf 3}, 623 (2007). 
\bibitem{band_dispersion}
T. Ohta \textit{et al}., Phys. Rev. Lett. {\bf 98}, 206802 (2007).
\bibitem{ando}
T. Ando, J. Phys. Soc. Jpn. {\bf 53}, 3101 (1984). 
\bibitem{hashimoto}
K. Hashimoto \textit{et al}., Phys. Rev. Lett. {\bf 101}, 256802 (2008).
\end{thebibliography}

\end{document}